\documentclass[12pt]{article}
\usepackage{amsmath,amssymb,amscd,latexsym,euscript,mathrsfs,textcomp}
\usepackage[all]{xy}
\usepackage[cp1251]{inputenc}

\newcommand{\NN}{{\,\mathbb N}}

\newcommand{\RR}{{\,\mathbb R}}

\newcommand{\mA}{{\mathcal A}}
\newcommand{\mI}{{\mathcal I}}

\newtheorem{theorem}{\bf Theorem}[section]

\newtheorem{proposition}[theorem]{\bf Proposition}

\newtheorem{definition}{\sc Definition}[section]
\newtheorem{example}[definition]{\sc Example}
\newtheorem{remark}[definition]{\sc Remark}

\def\leq{\leqslant}
\def\geq{\geqslant}

\begin{document}


\begin{center}
 {\large Generalized Asynchronous Systems}
\footnote{The paper is performed under the program of strategic development of state educational institution of higher professional education, \textnumero 2011-PR-054}
\end{center}
\centerline{A. A. Husainov, husainov51@yandex.ru}
\centerline{E. S. Kudryashova, ekatt@inbox.ru}

\begin{abstract}
The paper is devoted to a mathematical model of concurrency 
the special case of which is asynchronous system.
 Distributed asynchronous automata are introduced here.
It is proved that the Petri nets and transition systems with independence
can be considered like the distributed asynchronous automata. Time distributed
asynchronous automata are defined in standard way by the map which assigns time intervals to
events. It is proved that the time distributed
asynchronous automata are generalized the time Petri nets and asynchronous systems.
\end{abstract}


2000 Mathematics Subject Classification 68Q10, 68Q85

\section*{Introduction}

Time Petri nets \cite{pok2010}-\cite{pen1}, time event structures \cite{vir2008}, 
time transition systems \cite{hen1}, time transition systems with independence 
\cite{dou2008} are applied for studying of concurrent processes behavior in verification tasks.
They also are applied for software creating \cite{K2011}.
There are the tasks for which solution need a more general time models 
in spite of the fact that the Petri nets are very convenient models 
for concurrent computing systems (see for example \cite{X2011}).
Obvious generalization of time Petri nets for asynchronous systems, 
in which each transition is associated with the time interval, is not suitable to solve this problem.
The generalization of asynchronous systems, which allows to define time systems, is introduced in this paper.

\section{ Distributed asynchronous automata}

\begin{figure}[h]
$$
\xymatrix{
& s_1 \ar[rd]^{a_2}\\
s \ar[ru]^{a_1} \ar@{-->}[rd]_{a_2} && s'\\
& s_2 \ar@{-->}[ru]_{a_1}
}
$$
\caption{Axiom (ii) for asynchronous automata}\label{condas}
\end{figure}
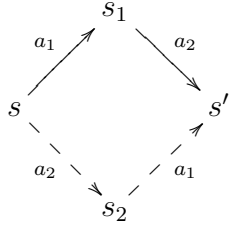

\begin{definition}

Distributed asynchronous automaton is a quintuple
$$
\mA = (S, s_0, E, \mI ,Tran)
$$
consisting of sets $S$ and $E$, element $s_0\in S$, relation
$Tran\subseteq S\times E\times S$ and the set of irreflexive symmetric
relations $\mI= (I_s)_{s\in S}$, $I_s\subseteq E\times E$. Following conditions must be satisfied

 (i) $(s,a,s')\in Tran ~\&~ (s,a, s'')\in Tran \Rightarrow s'=s'' $;

 (ii) for all $s\in S$, $(a_1,a_2)\in I_s$, $(s,a_1,s_1)\in Tran$ and
  $(s_1,a_2,s')\in Tran$ there is such $s_2\in S$ that $(s,a_2,s_2)\in Tran$
  and $(s_2,a_1,s')\in Tran$ (see fig.\ref{condas}).
\end{definition}

\begin{example}
Any asynchronous system $(S,s_0,E,I,Tran)$ can be considered as
distributed asynchronous automaton assumed that $I_s= I$ for all $s\in S$.
\end{example}

Definition of the automaton with independence was introduced in Goubault's paper \cite[Definition 3]{gou2002}. In the paper \cite{gou2010} interesting relations of this model with the Petri nets were established.

Goubault's definition differs from given above one so that the condition (ii) is replaced by following:

(ii)' For all $(a_1,a_2)\in I_s$ there exists $s_1, s_2, s'\in S$, for which
$(s,a_1,s_1)\in Tran$, $(s_1,a_2,s')\in Tran$, $(s,a_2,s_2)\in Tran$
  and $(s_2,a_1,s')\in Tran$ (see fig.\ref{condas}).

 Example of asynchronous system $S=\{s_0, s_1, s_2\}$, $E=\{a_1, a_2\}$, $I=\{(a_1,a_2), (a_2, a_1)\}$ with transitions
$$
	
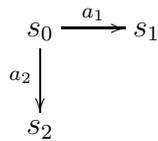
\xymatrix{
		s_0 \ar[d]_{a_2} \ar[r]^{a_1} & s_1\\
		s_2
	}
$$
shows that not any asynchronous system can be automation with independence.
Therefore Goubault's definition isn't more wide than ours. Moreover following statement, showing that any automaton with independence is distributed asynchronous automaton, is true.
\begin{theorem}
Any automaton $(S,s_0,E,\mI, Tran)$ with independence satisfies the axioms (i)-(ii) and so it is 
distributed asynchronous automaton.
\end{theorem}
{\sc Proof.}
Let $(S,s_0,E,\mI, Tran)$ satisfies to the
conditions (i) and (ii)'.
We will prove (ii). We consider $s\in S$ and couple $(a_1, a_2)\in I_s$ for this aim.
Let $(s, a_1, s_1)\in Tran$ and $(s_1, a_2, s')\in Tran$.
There are $r_1, r_2, r'\in S$ on account of (ii)' for which
$(s,a_1,r_1)\in Tran$, $(r_1,a_2,r')\in Tran$, $(r,a_2,r_2)\in Tran$
  and $(r_2,a_2,r')\in Tran$.

$$
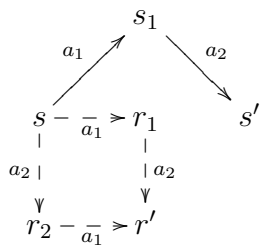
\xymatrix{
& s_1 \ar[rd]^{a_2}\\
s \ar[ru]^{a_1}
 \ar@{-->}[d]_{a_2} \ar@{-->}[r]_{a_1} & r_1 \ar@{-->}[d]^{a_2} & s'\\
r_2 \ar@{-->}[r]_{a_1} & r'
}
$$

On account of condition (i) it will be $r_1=s_1$ and $r'=s'$.
This implies the existence of transitions
$(s, a_2, r_2)\in Tran$ and $(r_2, a_1,s')\in Tran$.
\hfill $\Box$

\section{Petri nets as distributed asynchronous automata}

Petri net is a quintuple $(P,T, pre, post, M_0)$, consisting of finite sets $P$ and $T$, 
functions $M_0: P\to \NN$,
$pre: T\to \NN^P$, $post: T\to \NN^P$. At this point $\NN^P$ is a set of all
functions $P\to \NN$. The elements $p\in P$ are called {\em places}, ${t}\in {T}$ -- {\em transitions},
$M\in \NN^P$ -- {\em markings}, and $M_0$ -- {\em initial marking}.
We define the order relation on $\NN^P$ assumed that $M\leq M'$ if $M(p)\leq M'(p)$ is true for all $p\in P$.
We define amount and difference of functions as $(M \pm M') (p)= M(p)\pm M'(p)$.
For $M,M'\in \NN^P$ and $t\in T$ notation $M\stackrel{t}\to M'$ denotes that following two conditions are executed

\begin{enumerate}
\item $M \geq pre(t)$;
\item $M' = M - pre(t)+ post(t)$.
\end{enumerate}

In this case we speak that marking $M'$ is got from $M$ by transition $t$ firing.

Let $(P,T, pre, post, M_0)$ -- is Petri net.
We denote ${~}^{\bullet}t=\{p\in P: pre(t)(p)\not=0\}$.
For arbitrary marking $M\in \NN^P$ we define the relation
\begin{multline}\label{indep}
I_M=\{(t_1,t_2)\in T\times T: \\
M \geq pre(t_1) ~\& ~ M\geq pre(t_2)
\\
~ \& ~ {~}^\bullet{t_1} \cap {~}^\bullet{t_2}=\emptyset\}.
\end{multline}

\begin{theorem}
Any Petri net $(P,T, pre, post, M_0)$ defines a distributed asynchronous automaton  $(S, s_0, E, \mI, Tran)$
with $S=\NN^P$, $E=T$, $s_0=M_0$,
$Tran=\{(M, t, M')\in \NN^P\times T\times \NN^P : \mbox{ there exists }M\stackrel{t}\to M'\}$,
for which $I_M$ is denoted by formula (\ref{indep}).
\end{theorem}
{\sc Proof.}
If $(t_1,t_2)\in I_M$, then exist $M\stackrel{t_1}\to M_1$ and $M\stackrel{t_2}\to M_2$.
Therefore it is enough to show that for transitions firing
$$
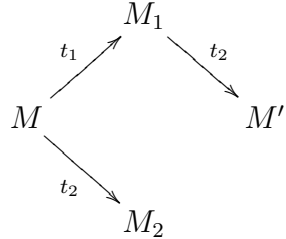
\xymatrix{
 & M_1 \ar[rd]^{t_2} \\
M\ar[ru]^{t_1} \ar[rd]_{t_2} && M'\\
& M_2
}
$$
$M_2\stackrel{t_1}\to M'$ will take place.
As the transition $t_2$ hasn't influence on counters which are located in entrance places of transition $t_1$ then $M_2\geq pre(t_1)$.
It have a place $M_2-pre(t_1)+post(t_1)= M- pre(t_2)+post(t_2) -pre(t_1)+post(t_1) =
M_1 - pre(t_2)+post(t_2) = M'$. So $M_2\stackrel{t_1}\to M'$.
\hfill $\Box$

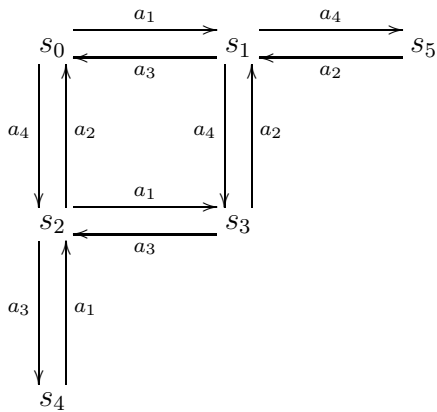
\begin{figure}[h]
$$
\xymatrix{
s_0 \ar@<1ex>[rr]^{a_1} \ar@<-1ex>[dd]_{a_4} && s_1 \ar@<1ex>[rr]^{a_4} \ar@<1ex>[ll]^{a_3}\ar@<-1ex>[dd]_{a_4}
			 && s_5 \ar@<1ex>[ll]^{a_2}\\
\\
s_2 \ar@<1ex>[rr]^{a_1} \ar@<-1ex>[dd]_{a_3} \ar@<-1ex>[uu]_{a_2}
		 && s_3  \ar@<1ex>[ll]^{a_3} \ar@<-1ex>[uu]_{a_2} \\
\\
s_4 \ar@<-1ex>[uu]_{a_1}
}
$$
\caption{Distributed asynchronous automaton for Petri net $\Omega$}\label{pic3}
\end{figure}

As an example we consider the following Petri net. It denotes by $\Omega$:

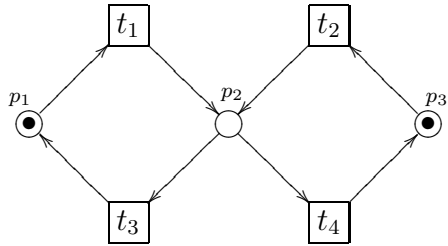
$$
\xymatrix{
& *+[F]{t_1} \ar[rd]^(0.85){p_2} & & *+[F]{t_2} \ar[ld]\\
*+[o][F]{\bullet} \ar[ru]^(0.1){p_1} && *+<9pt>[o][F]{~}\ar[ld] \ar[rd] && *+[o][F]{\bullet} \ar[lu]_(0.1){p_3}\\
& *+[F]{t_3}\ar[lu] & & *+[F]{t_4}\ar[ru]
}
$$
The set of reachable markings consist of
$M_0(1,0,1)$,  $M_1(0,1,1)$, $M_2(1,1,0)$,
$M_3(0,2,0)$, $M_4(2,0,0)$, $M_5(0,0,2)$.
Distributed asynchronous automaton, which is denoted by this Petri net, is shown on fig. \ref{pic3}.

We see that the states of distributed asynchronous automaton $s_i$, $0\leq i\leq 5$ correspond to
the Petri net markings $M_i$ and the actions $a_i$, $i \in \{1,4\}$ correspond to the transitions $t_i$.
The relations $I_s$ for this automaton are $I_{s_0}=\{(a_1, a_4),(a_4, a_1)\}$,
 $I_{s_1}=\emptyset$, $I_{s_2}=\{(a_1, a_2),(a_2, a_1)\}$, $I_{s_3}=\{(a_2, a_3),(a_3, a_2)\}$,  $I_{s_4}=\emptyset$,  $I_{s_5}=\emptyset$.

\begin{remark}
If we consider Petri net $\Omega$ as elementary then we will receive distributed asynchronous automaton which doesn't correspond to Goubault's definition \cite[Definition 3]{gou2002}.
\end{remark}

\section{Time distributed asynchronous automata}

We generalize definition of time Petri net is given in paper \cite{pop2010}.
We define as $\RR_{\geq 0}$ the set of all nonnegative real numbers.

\begin{definition}
Time distributed asynchronous automaton $(\mA, eft, lft)$ is a distributed asynchronous automaton
$$
	\mA= (S, s_0, E, \mI, Tran),
$$
with a couple of functions $eft: E\to \RR_{\geq 0}$, $lft: E\to \RR_{\geq 0} \cup \{\infty\} $
which satisfy to inequality $eft(a)\leq lft(a)$ for all $a\in E$.
\end{definition}

We introduce time states. We define reflection $S\times E\stackrel{\cdot}\to S\sqcup \{*\}$ assumed that $s\cdot a= s'$ if $(s,a,s')\in Tran$. If there aren't such  $s'\in S$  then assume $s\cdot a=*$.

\begin{definition}
Time state of time distributed automaton $(\mA, eft, lft)$ is a couple $(s,h)$ consisting of $s\in S$ and function $h: E\to \RR_{\geq 0} \cup \{\#\}$, such that
\begin{enumerate}
\item $s\cdot a \in S \Rightarrow h(a) \leq lft(a)$;
\item $s\cdot a =* \Rightarrow h(a)=\#$.
\end{enumerate}
\end{definition}

Each action $a\in E$ has a "clock". At the beginning of work time state equal to $(s_0, h_0)$ where
$h_0(a)=0$ if $s'\in S$ and transition $s\stackrel{a}\to s'$ exist.

\begin{definition}
We will write $(s,h) \stackrel{a}\to (s',h')$ and say
that action $a\in E$ transfers time state $(s,h)$ to $(s',h')$, if

(1) $s\cdot a= s' \not= * ~\&~ eft(a) \leq h(a)$;

(2) $(\forall b\in E)~ h'(b)=\left\{
\begin{array}{ll}
\# & \mbox{ if } s'\cdot b=*\\
h(b) & \mbox{ if and only if }  s'\cdot b\not=* ~\&~ (a,b)\in I_s\\
0 & \mbox{in the other case}.
\end{array}
\right.$
\end{definition}

\begin{definition}
For $\tau\in \RR_{\geq 0}$ we will write $(s, h)\stackrel{\tau}\to (s',h')$
and say, that state $(s,h)$ is replaced by state $(s',h')$
after the time $\tau$ is running out

(1) $s'=s$;

(2) $(\forall a\in E)~ h(a)\not=\# \Rightarrow h(a)+\tau \leq lft(a)$;

(3) $(\forall a\in E)~ h'(a)=\left\{
\begin{array}{ll}
\# & \mbox{ if } s'\cdot a=*\\
h(a)+\tau & \mbox{ if }  s'\cdot a\not=* .
\end{array}
\right.$

\end{definition}

\begin{proposition}
Definitions 3.2--3.4 generalize definition of time state and its modifications introducing for Petri nets in the paper \cite{pop2010}.
\end{proposition}

For example we consider asynchronous system consisting of two independent actions
$a_1$ and $a_2$ and four states
$$
	
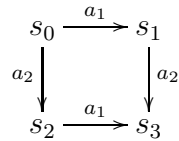
\xymatrix{
	 s_0\ar[d]_{a_2} \ar[r]^{a_1} & s_1\ar[d]^{a_2}\\
	 s_2 \ar[r]^{a_1} & s_3
	}
$$
for which $eft(a_i)$ and $lft(a_i)$, $i\in \{1,2\}$ are known.
We compute minimal time of operations performing which lead to state $s_3$.
We will consider time states $(s,h)$ as triplets $(s_i, \tau_1, \tau_2)$.
Let $eft(a_1)\leq eft(a_2)$.
Then following performing way can be
\begin{multline}
(s_0, 0, 0) \stackrel{eft(a_1)}\to (s_0, eft(a_1), eft(a_1)) \stackrel{a_1}\to (s_1, \#, eft(a_1))\\
\stackrel{eft(a_2)-eft(a_1)}\to (s_1, \#, eft(a_2)) \stackrel{a_2}\to (s_3, \#, \#)
\end{multline}
It is easy to see that obtained time equaling amount $eft(a_1)+eft(a_2)-eft(a_1)$ is minimal.
So in general case minimal time equals to $max(eft(a_1),eft(a_2))$.

We compute maximal time assume that $lft(a_1)\leq lft(a_2)$.
\begin{multline}
(s_0, 0, 0) \stackrel{lft(a_1)}\to (s_0, lft(a_1), lft(a_1)) \stackrel{a_1}\to (s_1, \#, lft(a_1))\\
\stackrel{lft(a_2)-lft(a_1)}\to (s_1, \#, lft(a_2)) \stackrel{a_2}\to (s_3, \#, \#)
\end{multline}

We obtain maximal time of action performance $max(lft(a_1),lft(a_2))$.

\section*{Conclusion}
Distributed asynchronous automata were introduced in the paper.
It permits to generalize time Petri nets on asynchronous systems and automata with independence.
Definitions of time states and occurrence actions on this states generalizing corresponding definitions for Petri nets were introduced.

\end{document}